\documentclass[useAMS,usenatbib,aas_macros]{mn2e}

\usepackage{hhline}
\usepackage{lscape}
\usepackage{graphicx}
\usepackage{colortbl}
\usepackage{amssymb}
\usepackage{amsmath}
\usepackage{natbib}
\usepackage{times}
\usepackage{aas_macros}
\usepackage{threeparttable}

\newcommand{\msun}{~\mathrm{M}_{\odot}}
\newcommand{\zsun}{~\mathrm{Z}_{\odot}}
\usepackage{color}

\def\simpropto{\lower.2ex\hbox{$\; \buildrel \propto \over \sim \;$}}
\def\ltsim{\lower.5ex\hbox{$\; \buildrel < \over \sim \;$}}
\def\gtsim{\lower.5ex\hbox{$\; \buildrel > \over \sim \;$}}

\begin{document}
\title[New constraints on DCBH formation]{New constraints on direct collapse black hole formation in the early Universe}
\author[B. Agarwal, et al.]{Bhaskar Agarwal$^{1,2}$\thanks{E-mail:
bhaskar.agarwal@yale.edu}, Britton Smith$^{3}$, Simon Glover$^4$, Priyamvada Natarajan$^{1}$, 
\newauthor{Sadegh Khochfar$^{3,2}$}\\
$^1$Department of Astronomy, 52 Hillhouse Avenue, Steinbach Hall, Yale University, New Haven, CT 06511, USA\\
$^2$Max-Planck-Institut f{\"u}r extraterrestrische Physik,
Giessenbachstra\ss{}e, 85748 Garching, Germany\\
$^3$Institute for Astronomy, University of Edinburgh, Royal Observatory, Edinburgh, EH9 3HJ\\
$^4$ Universit\"{a}t Heidelberg, Zentrum f\"{u}r Astronomie, Institut f\"{u}r Theoretische Astrophysik, Albert-Ueberle-Stra{\ss}e 2, 69120 Heidelberg, Germany}


\date{00 Jun 2014}
\pagerange{\pageref{firstpage}--\pageref{lastpage}} \pubyear{0000}
\maketitle

\label{firstpage}

\begin{abstract}
Direct collapse black holes (DCBH) have been proposed as a solution to the challenge of assembling supermassive black holes 
by $z>6$ to explain the bright quasars observed at this epoch. 
The formation of a DCBH seed with $\rm M_{BH}\sim10^{4-5}\msun$ requires a pristine atomic-cooling halo to be illuminated by an external radiation field that is sufficiently strong to entirely suppress H$_{2}$ cooling in the halo. Many previous studies have attempted to constrain the critical specific
intensity that is likely required to suppress H$_{2}$ cooling, denoted as $J_{\rm crit}$. However, these studies have typically assumed that the incident external radiation field can be modeled with a black-body spectrum. Under this assumption, it is possible to derive a {unique} value for $J_{\rm crit}$ that depends only on the temperature of the black-body. 
In this study we consider a more realistic spectral energy distribution (SED) for the external source of radiation that depends entirely on its star formation history and age.
The rate of destruction of the species responsible for suppressing molecular hydrogen cooling depends on the detailed shape of the SED. Therefore the value of $J_{\rm crit}$ is tied to the shape of the incident SED of the neighbouring galaxy. We fit a parametric form to the rates of destruction of H$_2$ and H$^-$ that permit direct collapse. Owing to this, we find that $J_{\rm crit}$ is not a fixed threshold but can lie anywhere in the range $J_{\rm crit} \sim 0.5$--$10^{3}$, depending on the details of the source stellar population.
 \end{abstract}

\begin{keywords}
quasars: general, supermassive black holes -- cosmology: darkages, reionization, firststars -- galaxies: high-redshift
\end{keywords}


\section{Introduction}

Current models of black hole formation and evolution are strongly challenged when it comes to explaining the existence of the population of observed bright {$z > 6$} quasars \citep{Mortlock:2011p447,2015ApJ...801L..11V,2015Natur.518..512W} that are believed to be powered by supermassive black holes (SMBHs) with $M_{\rm BH} \approx 10^9 \msun$. Growing these SMBHs from the remnant black holes produced by the first generation Population III (Pop III) stars poses a timing problem, as this needs to be accomplished within the first Gyr after the Big Bang. Pop III remnant masses are predicted to be low \citet[see e.g.][]{clark11,Stacy:2012p2734}, producing seed black hole (BH) masses  $M_{\rm BH} \sim 10-100 \msun$. {To grow these seeds by accretion into the observed supermassive black holes requires that they accrete with a low radiative efficiency at close to the Eddington rate for $\approx$ 800 Myr. This naturally requires the persistent presence of a gas reservoir to allow steady accretion. Additionally proper treatment of feedback from this central accreting BH must also be accounted for \citep{Alvarez:2009p778}. Recent theoretical work by \citet{Alexander14} has demonstrated that SMBHs at $z=6$ could indeed form via light BH seeds by super-boosting the growth of the seeds in gas-rich nuclear star clusters where the Eddington limit could in principle be circumvented.

{A promising way of avoiding the challenges faced by the light seeds invokes a different scenario} wherein seed BHs form with mass $M_{\rm BH} \approx 10^{4-5} \msun$ from direct collapse of primordial gas as proposed by \citep[see ][]{Eisenstein:1995p870,Oh:2002p836,Bromm:2003p22,Koushiappas:2004p871,Lodato:2006p375}. 
In order to bypass the formation of Pop III stars and make {direct collapse black hole} (DCBH) seeds, cooling and fragmentation of gas need to be thwarted in early collapsed dark matter halos. {The physical conditions that allow the formation of DCBHs require that} no coolants other than atomic hydrogen are available in these haloes \citep{Volonteri:2010p30, Natarajan:2011p90,Haiman12}. Such massive DCBH seeds are therefore theorised to form in pristine, metal-free atomic-cooling haloes where molecular hydrogen (H$_2$) formation can be suppressed. This occurs due to the presence of a critical level of {radiation in the Lyman-Werner
(LW) bands of H$_{2}$} in the energy range $11.2-13.6$~eV,
produced by external sources in the vicinity of the halo where the DCBH forms \citep[BA14 hereafter]{Dijkstra:2008p45,Agarwal14}.
Molecular hydrogen can cool the gas down to $\sim 100$~K, {leading} to fragmentation since the Jeans mass of such a cloud at a density $n=10^3\rm \ cm^{-3}$ {and temperature} $T = 100 \ {\rm K}$ is $\sim 1000 \msun$. The suppression of H$_2$ {formation} prevents the gas from cooling and forming Pop III stars, as the only available coolant in its absence is atomic hydrogen that can cool to $\sim 8000$ K. The Jeans mass of such a gas cloud at a density\footnote{At densities higher than $\sim 10^{3}\ \rm cm^{-3}$, the primary channel for H$_2$ destruction is collisional dissociation (KO01, CS10)} of $n=10^3\rm \ cm^{-3}$ {and temperature} $T = 8000 \ {\rm K}$ is $\sim 10^6 \msun$. Runaway collapse of this gas can lead to the formation of a seed BH in the nuclear regions that can retain up to 90\% of the Jeans mass \citep[see e.g.][]{Begelman:2006p3700,Regan:2009p776,Latif:2013p3629,2015MNRAS.450.4411C}. Most relevant to the final assembly of DCBHs are then the chemical pathways that involve the formation and destruction of molecular hydrogen in their host haloes. 

The two most critical chemical reactions that control the H$_2$ fraction in the collapsing gas are the photodissociation of H$_2$ and photodetachment of H$^-$
\begin{eqnarray}
&\rm H_2& +\ \gamma_{\rm LW} \rightarrow \rm H + H \label{reac.pdi} \\
&\rm H^-& +\ \gamma_{0.76} \rightarrow \rm H + e^{-} \label{reac.pde}
\end{eqnarray}
where $\gamma_{\rm LW}$ and $\gamma_{0.76}$ represent the photons in the LW band and {photons with energy greater than} $0.76$ eV respectively. {Destruction} of H$^-$ is critical to the process of DCBH formation as at low densities ($n<10^3\rm \ cm^{-3}$), {most of the H$_{2}$ formed in the gas is produced through the following reactions
\citep[e.g.][]{Lepp:1984p3301,Lepp:2002p2445}}
\begin{eqnarray}
&\rm H^{ }& +\ {\rm e} \rightarrow \rm H^- + \gamma \label{reac.hmfromh} \\
&\rm H^-& +\ \rm H \rightarrow \rm H_2 + e^{-} \label{reac.h2fromhm}
\end{eqnarray}
Therefore, the value of the specific intensity of the extragalactic radiation field in the LW band (defined here as $J_{\rm LW}$, the specific intensity at 13.6~eV in units of $10^{-21} \: {\rm erg \: s^{-1} \: cm^{-2} \: sr^{-1} \: Hz^{-1}}$)\footnote{{Note that other definitions of $J_{\rm LW}$ can also be found in the literature that consider the specific intensity at the mid-point of the LW band or averaged over the band.}} is the {key quantity that controls} the efficacy of the DCBH formation process. 
{This critical value of $J_{\rm LW}$ that is required to suppress H$_{2}$ formation sufficiently to allow direct collapse to occur is commonly referred to as the critical, threshold value $J_{\rm crit}$. Many authors have previously attempted to determine $J_{\rm crit}$, with the majority of these studies assuming} that the irradiating external source can be approximated as a black-body with a surface temperature of either $10^5$ K (a T5 spectrum) that is assumed to be representative of Pop III stars, or $10^4$ K (a T4 spectrum), taken to be representative of Population II stars (\citealt{Omukai:2001p128,Shang:2010p33,WolcottGreen:2011p121}, hereafter referred to as KO01, CS10 and WG11, respectively). 

{However, as we explore in this work, the actual spectral energy distributions (SEDs) of the first generation of galaxies are not particularly well approximated as black-bodies, and recent studies have shown that using more realistic galactic SEDs can have a significant impact on the value of $J_{\rm crit}$ (see \citealt{Sugimura:2014p3946} and \citealt{Agarwal15a}, hereafter KS14 and BA15, respectively).
In addition, the value of $J_{\rm crit}$ is sensitive to the details of the chemical network used to model the gas: changes in the set of reactions included in the network or in the rate coefficients adopted for them can lead to variations of a factor of a few in $J_{\rm crit}$ (see \citeauthor{Glover15a}~2015 -- hereafter SG15a -- and Glover 2015b). In simulations performed using a T5 spectrum, $J_{\rm crit}$ is also highly sensitive to the method used to model H$_{2}$ self-shielding: studies using the original \citet[DB96 hereafter]{Draine:1996p2556} self-shielding function find values of $J_{\rm crit}$ that are an order of magnitude larger than those from studies using the modified version of the DB96 function introduced in WG11 \citep{Sugimura:2014p3946}. To highlight these differences and dependencies, we summarise the values of $J_{crit}$ reported in literature in Table~\ref{tab.jcrit}.}

\begin{table}
\caption{Summary of the one--zone models in {the literature, listing the values of $J_{\rm crit}$ derived for T4 \& T5 spectra and the self-shielding model adopted}.}
\begin{threeparttable}
\begin{tabular*}{\columnwidth}{@{\extracolsep{\fill}}l | lll}
\hline
Study&$J_{\rm crit}$ & $J_{\rm crit}$ & Self-shielding \\
 & (T4) & (T5) & \\
 \\ [-1.5ex] \hline \\ [-1.5ex]
CS10 & 39 & 1.2$\times10^4$ & DB96\\
KS14 & 25 & 1.4$\times10^4$ & DB96\\

KO01$^{[\rm a]}$&18.1&3040&WG11$^{[\rm b]}$\\
WG11 & - & 1400 & WG11\\
KS14 & -  & 1600 & WG11\\
SG15a & 18$^{[\rm c]}$ & 1630$^{[\rm d]}$ & WG11\\
This work&18.8&1736& WG11 \\  [1.5ex] \hline
\end{tabular*}
\label{tab.jcrit}
\begin{tablenotes}
\item[a] {Computed using the KO01 network, but with updated chemical rate coefficients; see SG15a, Table 5}
\item[b] WG11: modified version of DB96 
\item[c] Run 2 in SG15a 
\item[d] Run 5 in SG15a
\end{tablenotes}
\end{threeparttable}
\end{table}

\begin{table}
\caption{Summary of the stellar populations considered in this study.}
\begin{threeparttable}
\begin{tabular*}{\columnwidth}{@{\extracolsep{\fill}}l | cccc}
\hline
ISb / CSf & M$_*$ / SFR & t$_*$ & Z & IMF\\ 
 & \tiny{($\msun$)} / ($\msun \rm yr^{-1}$) & \tiny{(yr)} & \tiny{($\zsun$)}  \\ 
 \\ [-1.5ex] \hline \\ [-1.5ex]
IA / CA&$10^{5-10}$ / $0.01-100$ &$10^{6-9}$~yr &0.5 & Salpeter$^{\rm [a]}$\\
IB / CB&$10^{5-10}$ / $0.01-100$ &$10^{6-9}$~yr &0.02 & Kroupa$^{\rm [b]}$\\ [1.5ex] \hline
\end{tabular*}
\label{tab.SBmodels}
\begin{tablenotes}
\item[a] {slope$:2.35$, Mass intervals $: 1,100\msun$}
\item[b] {slope$:1.3,2.3$, Mass intervals $:0.1,0.5,100\msun$}
\end{tablenotes}
\end{threeparttable}
\end{table}


{In order to} accurately capture the role played by LW photons in the cooling of gas in these early galaxies, {it is therefore necessary to use a chemical model that includes all the reactions important for determining $J_{\rm crit}$  as we do here. We use the best available values for the chemical rate coefficients, and properly account for the fact that the high-redshift galaxies that produce the photons responsible for destroying H$_{2}$ and H$^{-}$ have SEDs that are not simple black-bodies. 
In this paper, we present the results of calculations that improve upon the prior simplifications reported in the literature. 
Our paper is organised as follows. We begin by briefly outlining our methodology in Section~\ref{methodology}. 
{We present our results in Section~\ref{results}, and follow this in Section~\ref{discussion} with a discussion of their implications for the efficiency and feasibility of DCBH formation in the early Universe.}

\section{Methodology}
\label{methodology}
In this work we use the one-zone module of the publicly available adaptive--mesh--refinement code \texttt{Enzo} to follow the evolution of gas to high densities. This one--zone calculation with the up--to--date chemical networks for gas cooling (SG15a) tracks the detailed gas collapse in primordial haloes. To delineate the potential sites of DCBH formation, we model the incident radiation from an external source --first galaxies in the vicinity-- using {\sc{starburst 99}} {\citep{Leitherer:1999p112}}.

\subsection{{Basic model}}

The one-zone model employed here follows that of
\citet{Omukai:2000p3282}.   We set the initial gas temperature to $20,000$~K, and initial gas density to $10^{-1}\ \rm cm^{-3}$. The gas cooling threshold at constant density, i.e. the lowest temperature the gas can cool to, is set to the H$_2$ cooling limit $\sim 100$~K. The specific {internal} energy is evolved as
\begin{equation} 
\frac{de}{dt}= -p \frac{d}{dt} \frac{1}{\rho} - {\Lambda},   
\label{eq:energy}  
\end{equation} 
where the pressure is given by
\begin{equation} 
p = \frac{\rho k T}{\mu m_{\rm H}},
\end{equation} 
the specific internal energy is
\begin{equation} 
e = \frac{1}{\gamma_{\rm ad} -1} \frac{k T}{\mu m_{\rm H}},
\label{eq:defen} 
\end{equation}
{$\gamma_{\rm ad}$ is the true adiabatic index} and $\Lambda$ is the rate of radiative cooling.  The cooling rate and
associated {chemical} network is evolved by the
\texttt{Grackle}\footnote{https://grackle.readthedocs.org/en/latest/} 
chemistry and cooling solver \citep{2014ApJS..211...19B,
  2014ApJS..210...14K}.  The \texttt{Grackle} machinery was originally
extracted from the chemistry and cooling network of \texttt{Enzo}, 
{but has been updated to make use of the best available chemical and cooling rate data and to ensure that the chemical network contains all of the reactions important for determining $J_{\rm crit}$. Details of these updates and the reactions included in the chemical network can be found in the Appendix.}

As in \citet{Omukai:2000p3282}, the density is evolved as
\begin{equation} 
\frac{d \rho}{dt} = \frac{\rho}{t_{\rm col}},
\end{equation} 
where the collapse time-scale, $t_{\rm col}$
\begin{equation} 
t_{\rm col} = \sqrt{\frac{3 \pi}{32 G \rho}},
\end{equation} 

{To account for the effects of H$_{2}$ self-shielding, we assume that the gas is shielded by an H$_{2}$ column density given by 
\begin{equation}
N_{\rm H_{2}} = n_{\rm H_{2}} \lambda_{\rm J},
\end{equation}
where $n_{\rm H_{2}}$ is the H$_{2}$ number density and $\lambda_{\rm J}$ is the Jeans length. Based on this value, we then calculate the reduction in the H$_{2}$ photo-dissociation rate using the self-shielding model of WG11. We do not account for any reduction in the H$_{2}$ photodissociation rate caused by the absorption of LW photons by the Lyman series lines of atomic hydrogen \citep{Haiman:1997p85}.

{To validate our one-zone model, we calculate the values of $J_{\rm crit}$ corresponding to illumination by external 10$^{4}$ K or 10$^{5}$ K black-body sources. As our chemical model includes all of the key reactions identified by SG15a and uses an up-to-date set of chemical rate coefficients, we would expect to recover very similar values for $J_{\rm crit}$. As shown in Table~\ref{tab.jcrit}, our results agree to within a few percent with those presented in SG15a.}

\subsection{{Determining the photo-dissociation (H$_2$) and photo-detachment (H$^-$) rates}}

Here, we outline our computation of the rate coefficients for H$_{2}$ photo-dissociation and H$^{-}$ photo-detachment for a more realistic SED. 
Following BA14 we define the dimensionless parameter, J$_{LW}$, as the specific intensity at 13.6 eV, normalised to 10$^{-21}$ erg/s/cm$^2$/Hz/sr, for a point source at a distance $d$ (in cm).

\begin{eqnarray}
\label{eq.J}
{{J}}_{LW} \equiv \frac{\rm L_{13.6}}{\pi\ 4\pi d^2}\ {\rm erg/s/cm^2/Hz/sr} \\ \nonumber
\times \frac{1}{10^{-21} {\rm\ erg/s/cm^2/Hz/sr}}. 
\end{eqnarray}

\noindent Where we define a normalised spectrum, L$_n$, as

\begin{equation}
{\rm L}_n = \frac{{\rm L}_\nu}{\rm L_{13.6}} \times 10^{-21}\ {\rm erg/s/cm^2/Hz/sr}
\label{eq.normsed}
\end{equation}

which is the ratio of the spectrum L$_\nu$ to its value at 13.6 eV, in the units of $10^{-21}\ {\rm erg/s/cm^2/Hz/sr}$.

{The rate coefficients for H$_{2}$ photodissociation and H$^{-}$ photodetachment can be written as}
\begin{eqnarray}
&\rm k_{di} = \kappa_{di}\beta J_{LW} \rm \label{eq.h2} \\
&\rm k_{de} = \kappa_{de}\alpha J_{LW} \rm \label{eq.hm}
\end{eqnarray}
where {$\kappa_{\rm de} \approx 10^{-10} \: {\rm s^{-1}}$ and $\kappa_{\rm di} \approx 10^{-12} \: {\rm s^{-1}}$ (see BA15 for more details)}. The dimensionaless parameters {$\alpha$ and $\beta$ encapsulate the dependence of the rates on the spectral shape of the incident radiation field. For $\alpha$, we have (KO01,BA15)}
\begin{equation}
\alpha = \frac{1}{\kappa_{de}} \int\limits_{\nu_{0.76}}^{\nu_{13.6}} \frac{4\pi {\rm{L}}_{n}}{h\nu} \sigma_{\nu} {\rm d}\nu,
\end{equation}
where L$_{n}$ is as defined previously, $\sigma_{\nu}$ (cm$^2$) is the photodetachment cross-section \citep[see e.g.][]{Wishart:1979p3403,John:1988p3398}, $\nu_{0.76}$ and $\nu_{13.6}$ correspond to the frequency limits (Hz) at 0.76 eV and 13.6 eV respectively. 

For $\beta$, we have instead the much simpler expression (KO01,BA15)}
\begin{equation}
\beta =\rm \frac{\int\limits^{\nu_{13.6}}_{\nu_{11.2}}L_{\nu}d\nu}{L_{13.6}\Delta \nu_{\rm LW}},
\label{eq.beta}
\end{equation}
where ${\nu_{11.2}}\ \&\ {\nu_{13.6}}$ denote the frequency limits (Hz) corresponding to 11.2 and 13.6 eV respectively (i.e. the LW band), and $\Delta \nu_{LW} = {\nu_{13.6}} - {\nu_{11.2}}$. An alternative definition of the parameter $\beta =\frac{L_{12.4}}{L_{13.6}}$, i.e the ratio of the SED at 12.4 ev and 13.6 eV, has also been used in previous works (e.g. KS14, and see \citealt{Abel:1997p3456} for more details). Our choice is motivated  from the fact that the luminosity at discrete energies produced by different stellar synthesis codes (e.g. \citealt{Leitherer:1999p112} vs \citealt{Bruzual:2003p3256}) can vary more significantly than the bolometric luminosity in a band. We note here, that both definitions are equally valid as long as the same convention is adopted throughout a calculation.

To determine the H$^{-}$ photodetachment and H$_{2}$ photodissociation rate coefficients, we therefore need to specify three numbers: $\alpha$, $\beta$, and $J_{\rm LW}$.
Our procedure for determining these values for atomic-cooling halos illuminated by radiation from an external stellar population is the same as reported in BA15 and is described in more detail in Section~\ref{results} below.}

\begin{figure}
\includegraphics[width=0.75\columnwidth,angle=90,trim={0cm 0cm 1cm 0.5cm},clip]{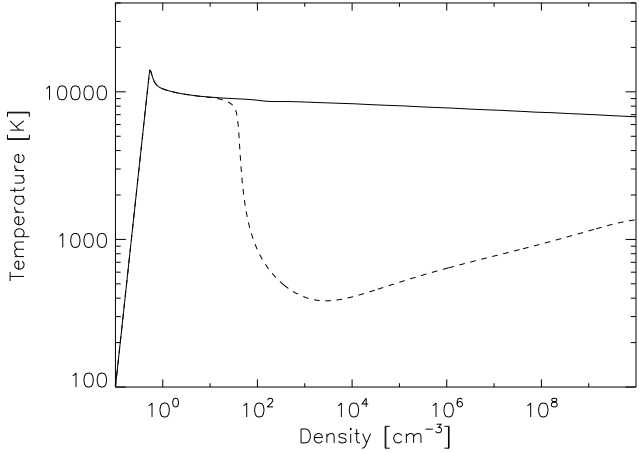}
\caption{Density-temperature plots for the collapse of the primordial gas in an atomic-cooling halo in the one-zone runs: DC (solid) and non-DC (dashed). The solid curve is produced when Eq.~\ref{eq.ratecurve} is satisfied by the values of $\rm k_{de}$ and $\rm k_{di}$ passed as inputs to the one-zone model, or in other words, when $J_{\rm LW} \gtsim J_{\rm crit}$. The dashed curve is produced when the gas is able to cool, i.e.\ the values of $\rm k_{di}$ and $\rm k_{di}$ values are not high enough and $J_{\rm LW} \ltsim J_{\rm crit}$.  }
\label{fig.rho_T}
\end{figure}

\section{Results}
\label{results}

We now present the results of the recomputation of the threshold intensity required for direct 
collapse with two key modifications compared to earlier work: and updated chemistry, and a more realistic assumption for the SED of
the irradiating source.

\subsection{A criterion for direct collapse}

We find that there is a region in the $\rm k_{de}$--$\rm k_{ di}$ parameter space where DCBH formation is permitted. To explore this allowed region we run our one-zone code for a large number of $\rm k_{de}$--$\rm k_{di}$ value sets, and identify those for which our one-zone model leads to an isothermal collapse at $\rm T=8000\ K$ with densities up to $n =10^{3} \rm\ cm^{-3}$ (see Fig.~\ref{fig.rho_T}), beyond which collisional dissociation of H$_2$ takes over (KO01, CS10). 
{We find that direct collapse occurs when $\rm k_{de}$ and $\rm k_{di}$ satisfy}
\begin{equation}
\rm k_{di} = 10^{Aexp(\frac{-z^2}{2}) + D}\ (\rm s^{-1}),
\label{eq.ratecurve}
\end{equation}
where $z=\frac{\log_{10}(\rm k_{de}) - B}{C}$ and $A = -3.864,\ B = -4.763,\ C = 0.773$, and $D = -8.154$, for $\rm k_{de}< 10^{-5}$.
{We plot the curve obtained from Eq.~\ref{eq.ratecurve} in Fig.~\ref{fig.ratecurve}. We see that for small $\rm k_{de}$ ($\ltsim 10^{-7} \ s^{-1}$), the value$\rm k_{di}$ of $\rm k_{di}$ required for direct collapse remains roughly constant at $\approx 7 \times 10^{-9}\ s^{-1}$. This implies that most of the H$^{-}$ ions that form in the gas are consumed by reaction \ref{reac.h2fromhm}, forming H$_{2}$. These H$_2$ molecules are destroyed via reaction \ref{reac.pdi} as $\rm k_{di}$ is high enough. 
Therefore as long as the production of H$_2$ resulting from the low H$^{-}$ photo-detachment rate can be countered by $\rm k_{di}$, changes in $\rm k_{de}$ have very little influence on the value of $\rm k_{di}$ required for direct collapse. 
However when $\rm k_{de}$ becomes large, it becomes the dominant destruction mechanism for H$^{-}$, thus rendering reaction \ref{reac.h2fromhm} ineffective. Once this occurs, further increases in $\rm k_{de}$ strongly suppress H$_{2}$ formation, thus allowing for direct collapse at smaller values of $\rm k_{di}$.}

{The diagonal lines in Fig.~\ref{fig.ratecurve} illustrate the relationship between $\rm k_{di}$ and $\rm k_{de}$ for a T4 (dotted line) and T5 spectrum (dashed line) respectively. These lines are defined by $\alpha = 2000$, $\beta = 3$ for the T4 spectrum and $\alpha = 0.1$, $\beta = 0.9$ for the T5 spectrum. The enormous difference in $\alpha$ for these two spectra 
demonstrates} the importance of taking into account the spectral shape of the irradiating source while computing the rate constants (BA15, KS15). {In particular, we see that for the T5 spectrum, the low H$^{-}$ photo-detachment rate implies that when the criterion for direct collapse is satisfied, photo-detachment remains chemically unimportant. On the other hand, for the T4 spectrum, H$^{-}$ photo-detachment is very important and strongly affects the required value of $\rm k_{di}$.}
It is therefore important to understand the effect of changing the irradiating source from a black-body to a {more} realistic SED, as the relative abundances of $\sim 1$~eV photons and LW photons change significantly as a function of the mass, star formation rate (SFR) and age {of the stellar population acting as the source for the radiation.}

\begin{figure}
\includegraphics[width=0.75\columnwidth,angle=90,trim={0cm 0cm 1cm 1cm},clip]{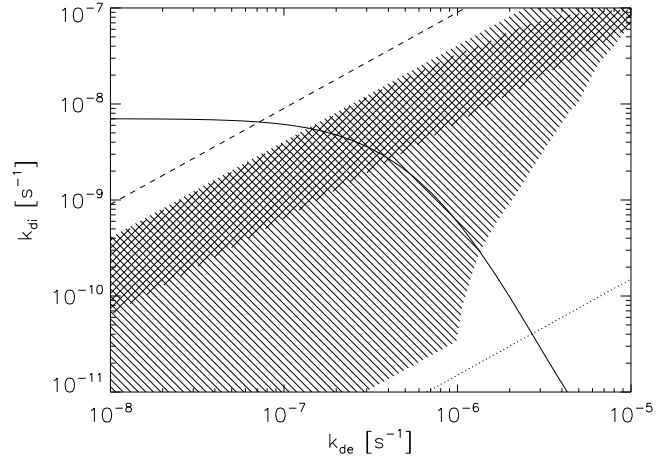}
\caption{The solid curve shows the criterion for direct collapse that we have derived from our one-zone runs (Eq.~\ref{eq.ratecurve}), while the diagonal lines show the combinations of $\rm k_{de}$ and $\rm k_{di}$ that we get for a T4 spectrum (dotted line) and a T5 spectrum (dashed line). For blackbody spectra, changing the value of $J_{\rm LW}$ changes our location along these diagonal lines, with $J_{\rm LW} = J_{\rm crit}$ at the point where the lines cross the solid curve. The shaded regions show the values of $\rm k_{de}$ and $\rm k_{di}$ produced by the stellar populations considered in Section~\ref{stellarpop}, computed for an assumed distance of 5~kpc. The region shaded with hatched--lines at 45$^\circ$ shows the results from the ISb model and the region shaded with hatched--lines at 135$^\circ$ shows the results from the CSf model. It is clear that neither of the commonly-adopted black-body spectra provide a good approximation of these shaded regions.
\label{fig.ratecurve}}
\end{figure}
\begin{figure*}
\centering
\includegraphics[width=0.75\columnwidth,angle=90,trim={0cm 1cm 1cm 1cm},clip]{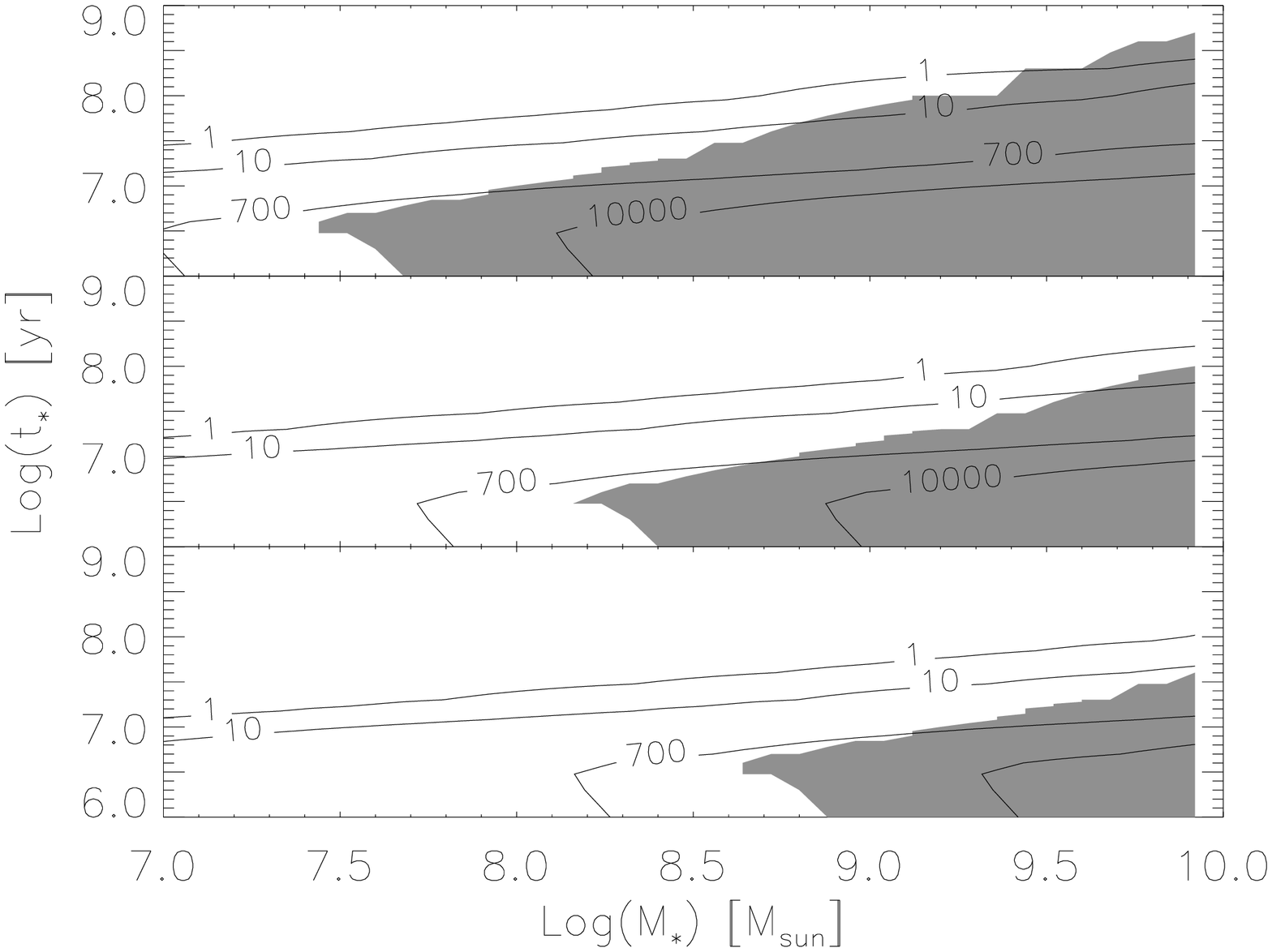}
\includegraphics[width=0.75\columnwidth,angle=90,trim={0cm 1cm 1cm 1cm},clip]{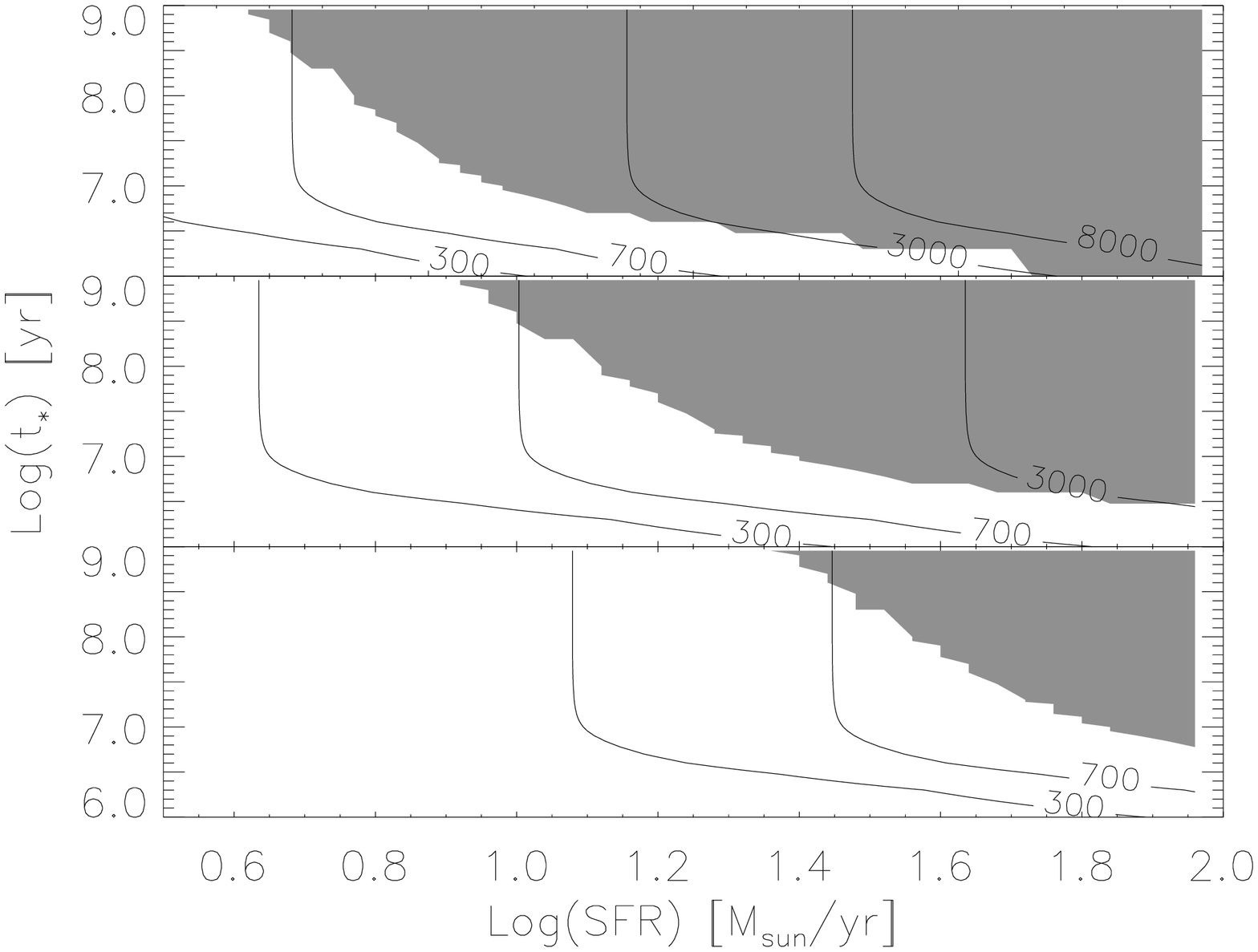}
\caption{Stellar populations that allow for DC. Left: {results for the ISb model, IA}; right: {results for the CSf model, CA}. The shaded regions in grey {indicate the stellar populations} that satisfy Eq. \ref{eq.ratecurve} {for an assumed separation} of 5, 12 and 20 kpc (top, middle and bottom panels, respectively) between the atomic cooling halo and the irradiating source. The contours of $J_{\rm LW}$ at the respective distances are over-plotted in each of the panels.}
\label{fig.conts}
\end{figure*}

\begin{figure*}
\centering
\includegraphics[width=0.7\columnwidth,angle=90,trim={-1cm 0cm 0cm -1cm},clip]{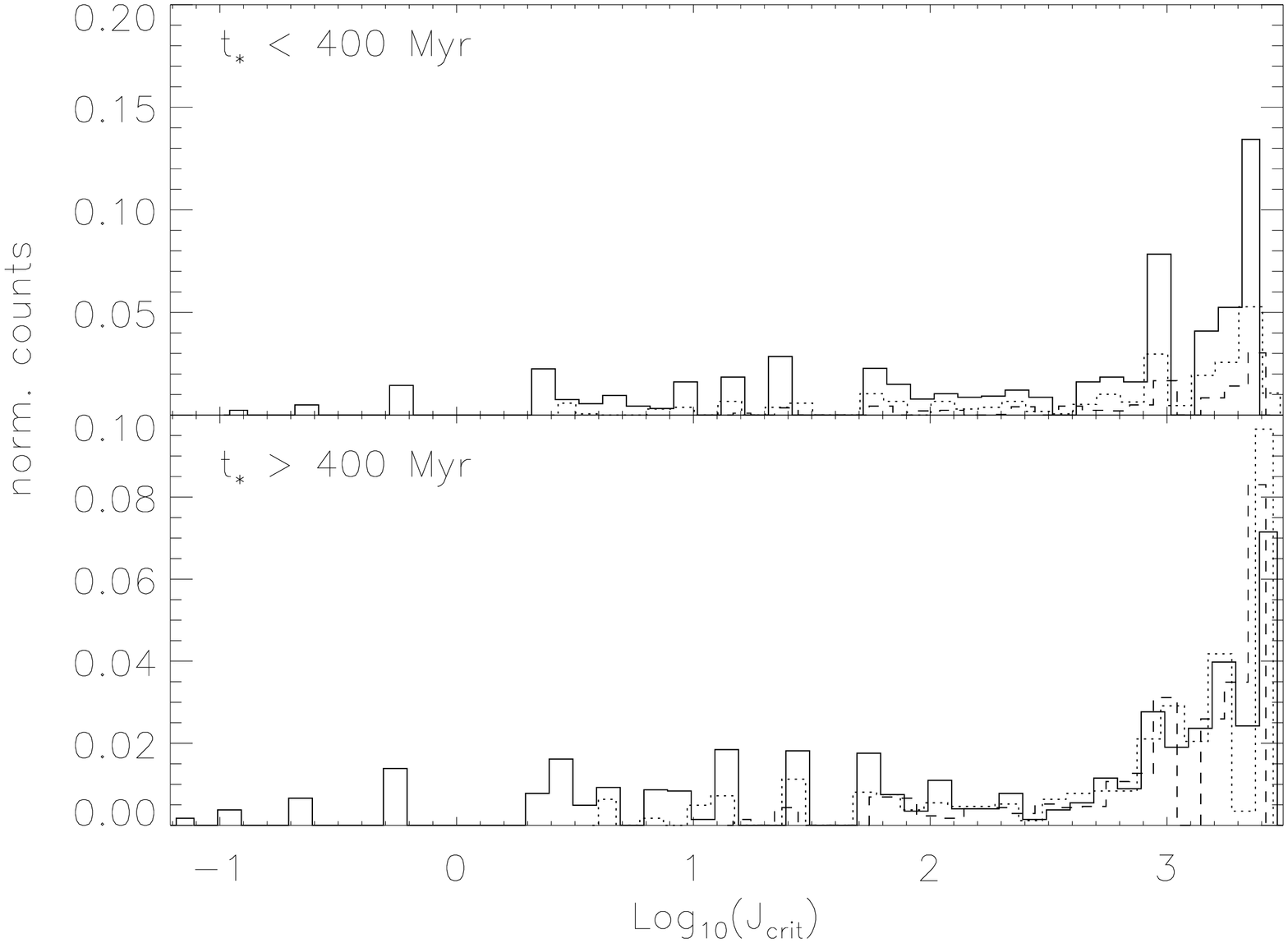}
\includegraphics[width=0.7\columnwidth,angle=90,trim={-1cm 0cm 0cm -1cm},clip]{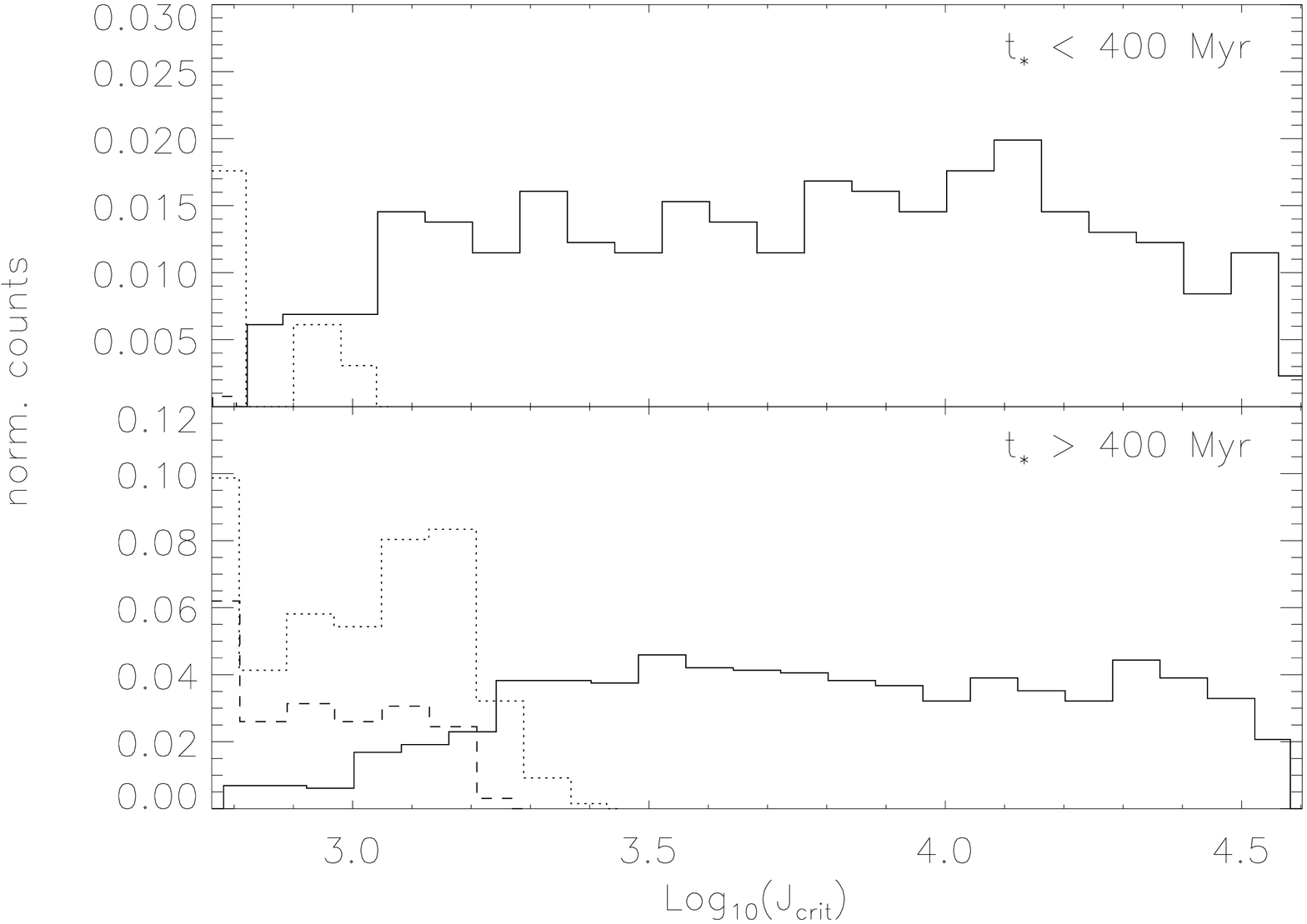}
\caption{{Histograms of $J_{\rm crit}$ for the ISb model, IA (left panel) and the CSf model, CA (right panel). {The histograms are plotted by splitting the stellar populations into ones with $t_*>400$~Myr and $t_*<400$~Myr}. The values of $J_{\rm crit}$ are obtained by requiring that Eq.~\ref{eq.ratecurve} be valid in the grey regions in Fig. \ref{fig.conts}. The solid, dotted and dashed lines correspond to the 5, 12 and 20 kpc separations respectively.}}
\label{fig.jhist}
\end{figure*}

\subsection{Stellar populations that can lead to DC in their vicinity}
\label{stellarpop}
With this understanding of the conditions that lead to direct collapse, we can now ask 
what sort of realistic stellar populations are able to give rise to {the appropriate combinations of $\rm k_{de}$ and $\rm k_{di}$ that satisfy the criterion for direct collapse?}
{To explore this, we create} a grid of stellar populations on the age--mass and age--SFR plane, where each point in the grid represents a unique SED. {Based on this grid of SEDs, we then calculate the corresponding values for $\rm k_{de}$ and $\rm k_{de}$ at various assumed distances between the atomic-cooling halo (DCBH host) and the stellar population.} We outline the steps {in our} approach below.

\begin{enumerate}
\item[{\bf Step 1}:] 
We use {\sc starburst99} \citep{Leitherer:1999p112} to generate individual SEDs (not including nebular emission) for an entire set of stellar populations 
at different stages in their evolution. {As in BA15, we consider two extreme cases for the star formation history: an instantaneous starburst model (ISb) where all the stars form instantly at $t = 0$, and a continuous star formation model (CSf) in which we assume that the star formation rate $\dot{M}_{*}$ remains constant over the lifetime of the stellar population.}
For each stellar population, we need to specify the metallicity, along with the total stellar mass $M_{*}$ (ISb) or the star formation rate (CSf), and $t_{*}$ which denotes the age of the stellar population (time since the onset of star formation) in case of ISb (CSf). The values we consider are summarised in Table \ref{tab.SBmodels}. In the ISb models, we generate 1800 and 3600 SED models for the IA and IB cases respectively. For the CSf case, we generate 15550 and 11400 SED models for the CA and CB case respectively. This produces a finer grid of SEDs (dependent on galactic properties) as compared to previous work by KS14 where they generated 64 and 208 models for their ISb and CSf cases respectively.


{We choose a narrow range in metallicity} as BA15 and KS14 {have} demonstrated that the {values of $\rm k_{de}$ and $\rm k_{di}$ are far} more sensitive to the choice of the age and mass (or SFR) than the metallicity of the external source.  
{Furthermore, we assume an escape fraction, $f_{esc} =1$, and an optically thin intergalactic medium. Changing these parameters will only reduce the value of $J_{LW}$ and should not qualitatively affect our results.}

\item[{\bf Step 2}:]
For a given SED, we then follow the methodology in BA15 to evaluate the rates (Eq. \ref{eq.h2} \& \ref{eq.hm}) by computing:
\begin{itemize}
\item $\alpha,\ \beta$,
\item $J_{\rm LW}$ is computed using Eq.~\ref{eq.J} {for assumed physical distances of 5, 12 and 20 kpc between the putative DCBH host halo and the irradiating source}. 
\end{itemize}
\end{enumerate}

{The values of $\rm k_{de}$ and $\rm k_{di}$ that we obtain using this procedure are illustrated in Fig.~\ref{fig.ratecurve}. The region shaded with hatched--lines at a 45$^\circ$ corresponds to the ISb model, while that shaded with hatched--lines at a 135$^\circ$ corresponds to the CSf model. In both cases, we assume that $d = 5$~kpc.\footnote{Note that the corresponding regions for $d = 12$ or 20 kpc can be computed easily, since both $\rm k_{di}$ and $\rm k_{de}$ scale as $d^{-2}$. They are omitted from the Fig. merely for clarity.} 
We see that both models produce a similar range of values for $\rm k_{de}$ since most of the photons contributing to $\rm k_{de}$ are produced by long-lived lower-mass stars. On the other hand, the UV photons responsible for H$_{2}$ photo-dissociation are produced primarily by high-mass, short-lived stars, and so the CSf model tends to produce much higher values for $\rm k_{di}$ than the ISb model. We find an overlap between the two regions only when the underlying SEDs are very similar, which typically occurs  only when $t_{*}$ is small. For instance,} a 10~Myr old stellar population with a mass $M_*=10^7 \msun$ that was produced in a single instantaneous burst {produces a very similar SED to one that} forms with a constant SFR of $\dot M_* = 1 \msun \: {\rm yr^{-1}}$ for the past 10 Myr.\footnote{Note, however, that the CSf model naturally produces a slightly bluer spectrum with increasing age, since there are more high mass stars at any given point in the stellar population's evolution than with the ISb model.} Although most of the stellar populations considered here lie below the curve, the reaction rates produced by the stellar models that allow for direct collapse span three orders of magnitude in both $\rm k_{de}$ and $\rm k_{di}$.
We also see from Fig.~\ref{fig.ratecurve} that the T4 and T5 spectra commonly used when studying direct collapse do not provide a good description of the values of $\rm k_{de}$ and $\rm k_{di}$ produced by realistic stellar populations.
An important question now arises: does the intersection of the curve and the shaded regions in Fig.~\ref{fig.ratecurve} lead to a single well-defined value of $J_{\rm crit}$?

\subsection{There is no unique $J_{\rm crit}$}
\label{sec.nojcrit}

In order to explore the implications of Fig. \ref{fig.ratecurve} {for the value of} $J_{\rm crit}$, {the properties of the stellar populations, i.e. $M_* {\ \rm or\ } \dot M_*\ \&\ t_*$, that produce photo-dissociation and photo-detachment rates large enough to enable direct collapse are show in Fig.~\ref{fig.conts}. {The left-panel shows the results for the IA SEDs (ISb), which are characterized by their age and stellar mass, while the right-panel shows results for the CA SEDs (CSf), which are characterized by their age and SFR. Qualitatively, IA, IB, CA and CB, lead to the same result, thus we only show the IA and CA cases here, and point the reader to the Appendix where additional results of IB and CB are shown.}   The three rows (top--bottom) in each panel correspond to separations of 5, 12 and 20 kpc {respectively. The grey shaded region demarcates the parameter space for which direct collapse is permitted. Finally, the contours correspond to the indicated values of $J_{\rm LW}$. We note that in every case, the boundary of the shaded region crosses more than one contour, indicating that it does not correspond to a single fixed value of $J_{\rm LW}$. In other words, there is no single value of $J_{crit}$ in either the ISb or CSf case.
We discuss the {behavior} of $J_{\rm crit}$ for our two {star formation models in more detail below}.

\paragraph*{{ISb}} \mbox{ }

\noindent At a distance of 5 kpc, young stellar populations with masses as low as $M_*\sim10^8 \msun$ and $t_*\sim5\ \rm Myr$ are able to satisfy Eq.~\ref{eq.ratecurve}, leading to conditions conducive for DCBH formation. 
{As we increase the separation, the viable ages decreases and the required stellar mass increases} to compensate for the $d^{-2}$ scaling of the {radiation field}, increasing by a factor of 5.76 and 16 for $d = 12$ and 20~kpc, respectively.

{As previously noted}, the shaded region that permits direct collapse is not bounded by a single $J_{\rm LW}$ contour. 
{Young and low stellar mass populations with $10^{7}\ltsim M_* \ltsim 10^{8.5} \msun$ and $1\ltsim t_*\ltsim 50$~Myr are represented by $J_{LW}\gtsim 700$. Stellar populations older than 50 Myr and with $M_* \gtsim 10^{8.5} \msun$ lie in the region where $J_{LW}\ltsim700$ indicating that in this case, much smaller values of $J_{\rm crit}$ would suffice. Indeed, the smallest value of $J_{\rm crit} \sim 0.1$ in our parameter space exploration is obtained when $M_* \sim 10^{10} \msun $, $t_* \sim 300\ \rm Myr$ and $d = 5$~kpc.}

A range of $J_{crit}$ values can be found by imposing Eq.~\ref{eq.ratecurve} on the $\rm k_{de}$--$\rm k_{di}$ values and systematically lowering the values of $J_{LW}$ in the contours above, till a minimum critical '$J_{crit}$' value is found for which the equation is still valid. Histograms of these $J_{crit}$ values are shown in Fig.~\ref{fig.jhist}.
{Assuming that DCBHs grow rapidly to $M_{BH}\approx 10^9 \msun$ by Eddington accretion with $f_{edd}=1$ from $z=12$ (redshift of DCBH seed formation) to $z=7$, we split the stellar populations with an age limit of 400 Myr, i.e. the age of the Universe at $z=12$.} 
We see that for a source located at 5~kpc, the value of $J_{\rm crit}$ ranges from $\sim 0.1 - 1000$ depending on the age and mass of the stellar population. For larger distances, we find less variation in $J_{\rm crit}$, although this largely just reflects the fact that fewer of the stellar populations in the parameter range that we consider produce enough radiation to cause direct collapse.

\paragraph*{CSf} \mbox{ }

\noindent {In the case of the continuous star formation model, there is less variation in the value of $J_{\rm crit}$, as can be seen from the right-hand panels of Fig.~\ref{fig.conts} and Fig.~\ref{fig.jhist}. Direct collapse is not permitted when $J_{\rm LW} \ltsim 700$, which might lead one to infer that $J_{\rm crit} \sim 700$. However, this is clearly misleading -- as there are large regions of parameter space (shown clearly in Fig. \ref{fig.conts}) where $J_{\rm LW} > 700$ for which direct collapse still remains impossible. Notably, the actual values of $J_{crit}$ vary by at least {an order of magnitude} as we change the age and the star formation rate of the stellar population illuminating our halo.} {As in the ISb case, we conclude that there is no absolute value of $J_{\rm crit}$ that we can determine independent of the properties of the illuminating source to assess the feasibility of direct collapse.}

{It is clear from Fig.~\ref{fig.conts} \& \ref{fig.jhist} that any attempt to describe this behaviour using a single value for $J_{\rm crit}$ is inadequate. The critical flux depends in a complicated fashion on the mass and age of the stellar population and its distance from the atomic-cooling halo. Depending on the value of a fixed $J_{crit}$ one may therefore either dramatically over-estimate or under-estimate the comoving number density of atomic-cooling halos that could host DCBHs. }

\section{Conclusions and Discussion}
\label{discussion}
The mean background value of $J_{\rm LW}$ in the early Universe is expected to rise with the cosmic SFR, reaching a value of $\sim 1$ at $z=10$ \cite[see e.g.\ KO01,][]{Greif:2006p99}. Current estimates of the critical value, $J_{\rm crit}$, are at least an order of magnitude higher than the mean background, which implies that DC sites will {preferentially be found} close to the first galaxies \citep[BA14]{Agarwal12}. However, any theoretical model that attempts to explore the abundance of such sites is extremely sensitive to the choice of $J_{\rm crit}$. {For example}, an order of magnitude increase in $J_{\rm crit}$ can reduce the DCBH number density at $z=6$ by three orders of magnitude \citep{Dijkstra:2008p45,2014MNRAS.443.1979L,Dijkstra2014a}. This {emphasizes} the need to {understand} and calculate the value of $J_{\rm crit}$ produced by galaxies composed of Pop II stars in the early Universe.

In this study, we have shown that a $J_{\rm crit}$ threshold is not the correct way to identify direct collapse haloes. {The quantities that determine whether pristine gas in an atomic-cooling halo can undergo direct collapse are the values of the H$_{2}$ photo-dissociation rate, $\rm k_{di}$, and the H$^{-}$ photo-detachment rate, $\rm k_{de}$.} 
{Traditionally, a simple blackbody or power-law spectrum has been used to model the radiation from galaxies, resulting in a fixed ratio of these rate coefficients.  This made it possible to identify direct collapse sites using a single value for $J_{\rm crit}$.  We have shown here that such an approach is inadequate when realistic SEDs are taken into account.  Instead, one must consider the precise age, mass (or SFR), and distance of the stellar population producing the radiation field in order to know whether direct collapse will occur at a given site.} {Failing to account for the detailed shape of the irradiating spectrum could lead to a severe {underestimation or overestimation} of the {comoving number density of} DC sites in the early Universe.

{Comparing our work to the previous study of KS14, we note a difference in approach. KS14 assume a fixed value for the rate parameter that goes into computing $\rm k_{di}$, while we derive the rate parameter for $\rm k_{di}$ directly from the spectrum itself.} This inevitably leads to a larger variation in $J_{\rm crit}$ (especially for the ISb case) than that reported by KS14. This could also be due to the larger range in stellar masses considered in this study as compared to KS14 (more details are provided in sec.~\ref{stellarpop}).

We expect these results to depend on the choice of IMF. For an extremely top--heavy IMF, the results may vary significantly since the stellar populations would be predominantly composed of {\textit{bluer}} stars, thereby pushing the $k_{di}$ to higher values, i.e. a regime where $\rm k_{di}$ is insignificant. {In this work we have neglected the contribution of nebular emission lines. These will only serve to add to the total radiation budget of the radiation sources.}
 
{The galaxies irradiating the pristine atomic cooling halo are also the source of metals that could in principle pollute the DCBH site. However in their cosmological hydrodynamical simulation, part of the FiBY project, BA14 found that the DCBH sites they identified in their volume were metal free and had not been polluted in their past by either galactic winds from neighbouring stellar populations, or in--situ Pop III star formation. additionally, studies have shown that metal mixing appears to be inefficient at these epochs \citep{Cen:2008p841,Britton15}.

So far we have discussed the possibility of DC in the vicinity of a single stellar population. However, the presence of multiple stellar populations could further aid DC, as the burden of producing the right {photo-dissociation and photo-detachment rates}  will then be shared by multiple stellar populations (BA14). The ideal case scenario would be the presence of an old and young stellar population nearby, which can give rise to optimal values of $\rm k_{de}$ and $\rm k_{di}$, respectively. {In this case, it makes even less sense to specify a single value for $J_{\rm crit}$, since the strength of the radiation flux required from each population depends not only on the SED of the population but also on how much radiation is produced by the other nearby source(s).} 

\section*{Acknowledgements}
BA would like to thank Zoltan Haiman and Kazu Omukai for discussions that prompted the birth of this idea.  BA would also like to thank Jarrett Johnson, Laura Morselli, Alessia Longobardi and Jonny Elliott for their useful comments on the manuscript.
SCOG acknowledges support from the Deutsche Forschungsgemeinschaft  via SFB 881, ``The Milky Way System'' (sub-projects B1, B2 and B8) and SPP 1573, ``Physics of the Interstellar Medium'' (grant number GL 668/2-1), and by the European Research Council under the European Community's Seventh Framework Programme (FP7/2007-2013) via the
ERC Advanced Grant STARLIGHT (project number 339177). PN acknowledges support from a NASA-NSF Theoretical and Computational Astrophysics Networks award number 1332858. BA acknowledges support of a TCAN postdoctoral fellowship at Yale. 
\bibliographystyle{mn2e}
\bibliography{babib}
\newpage
\appendix
\section[]{Chemical network}
{The chemical reactions used in this study are listed in Table~\ref{tab:reduce}. The chemical network is based on the original \texttt{Enzo} network described in \citet{2014ApJS..211...19B}, but we have added two new reactions and updated the rate coefficients used for seven others, as described below. This modified network includes all of the reactions in the minimal reduced model of SG15a that were identified as being crucial for an accurate determination of $J_{\rm crit}$.}

\subsection{New reactions}
{We have added two new chemical reactions to the primordial chemistry network implemented within \texttt{Grackle}: the collisional ionization of atomic hydrogen by collisions with hydrogen atoms (reaction 25) and with helium atoms (reaction 26). Reaction 25 was included in the study of KO01, but has been omitted in most subsequent studies of the direct collapse model. However, it is important to include this process as SG15a show that the additional ionization produced by this reaction has a significant effect on the value of $J_{\rm crit}$. Reaction 26 was not considered in any studies of the direct collapse model prior to SG15a but also proves to be important, albeit less so than reaction 25. We note that the rate coefficient for reaction 25 at the temperatures of interest is highly uncertain \citep{Glover15b}; we use the value of the rate coefficient given in \citet{Lenzuni91} for consistency with the study of KO01, but do not vouch for its accuracy.}

\subsection{Updated reaction rates}
\subsubsection*{H$^{-}$ formation by radiative association (reaction 7)} 
{The rate coefficient used for this reaction within \texttt{Enzo} is taken from \citet{hutchins76}, but his fit is valid only in the temperature range $100 < T < 2500$~K. We have therefore replaced it with the improved fit given in \citet{sld98} which is valid over a much broader range of temperatures.}

\subsubsection*{H$_2$ formation by associative detachment of H$^{-}$ (reaction 8)}
{The rate coefficient for this reaction has recently been measured in the temperature range $10 < T < 10^{4}$~K by \citet{kreckel10}. We use the analytical fit that they give to their experimentally-determined values.} 

\subsubsection*{H$_{2}^{+}$ formation by radiative association (reaction 9)}
{The rate coefficient previously used within \texttt{Enzo} and \texttt{Grackle} was an analytical fit given in \citet{1987ApJ...318...32S} and based on data from \citet{rp76}. However, this fit disagrees with the data on which it is based by up to 15\% at the temperatures of interest in the present study. We have therefore replaced it with the improved analytical fit given in \citet{Latif15}, which agrees with the \citet{rp76} data to within a few percent.} 

\subsubsection*{Collisional dissociation of H$_{2}$ by H (reaction 13)}
{The chemical model used in \texttt{Enzo} accounts for the contribution made to the rate of this reaction by direct collisional dissociation, but not the contribution coming from dissociative tunneling, which can dominate at low temperatures. \citet{2014MNRAS.443.1979L} and SG15a have shown that omitting this process introduces a factor of two uncertainty into estimates of $J_{\rm crit}$. We have therefore included it in our chemical model, using the rate given in \citet{martin96}.}  

\subsubsection*{Mutual neutralization of H$^{-}$ by H$^{+}$ (reaction 16)}
{\texttt{Enzo} uses a rate coefficient for this reaction that is taken from \citet{dl87}, but which disagrees by a factor of a few with other determinations of the rate at low temperatures \cite{Glover2006}. In our model, we use instead the rate coefficient given in \citet{croft99}, which agrees well with the results of more recent calculations and measurements of the mutual neutralization rate.}

\subsubsection*{H$^-$ photo-detachment (reaction 22)}
{Our treatment of this reaction is discussed in detail in Sections~\ref{methodology} and \ref{results}.}

\subsubsection*{H$_{2}^+$ photo-dissociation (reaction 23)}
{For this reaction, we follow the reasoning of KS14 and approximate the rate coefficient as
\begin{equation}
k_{\rm H_2^+} \approx 0.1  k_{\rm de}. \nonumber
\end{equation}
The justification for this approach is that in general, the integral of the cross-section times the incident spectrum for H$_2^+$ is around a factor of ten lower than for H$^{-}$, a result which is largely independent of the shape of the incident spectrum \citep[see][]{Sugimura:2014p3946}. In practice, the behaviour of the gas is not particularly sensitive to the rate of this reaction \citep[SG15a,][]{Glover15b}, and so this approximation does not significantly affect our results.}

\begin{table} 
\caption{List of reactions used in this study \label{tab:reduce}}
\begin{tabular}{clc}
\hline
No.\ & Reaction & Reference \\
\hline
1 & ${\rm H + e^{-}}  \rightarrow {\rm H^+ + e^- + e^-}$ & A97 \\
2 & ${\rm H^{+} + e^{-}} \rightarrow {\rm H + \gamma}$ & F92 (Case B)\\
3 & ${\rm He + e^{-}} \rightarrow {\rm He^{+} + e^{-} + e^{-}}$ & A97 \\
4 & ${\rm He^{+} + e^{-}} \rightarrow {\rm He + \gamma}$ & BS60 (Case B) \\
5 & ${\rm He^{+} + e^{-}} \rightarrow {\rm He^{++} + e^{-} + e^{-}}$ & A97 \\
6 & ${\rm He^{++} + e^{-}} \rightarrow {\rm He^{+} + \gamma}$ & F92 (Case B) \\
7$^{*}$ & ${\rm H + e^{-}} \rightarrow {\rm H^{-} + \gamma}$  & SLD98 \\
8$^{*}$ & ${\rm H^{-} + H} \rightarrow  {\rm H_{2} + e^{-}}$ & K10 \\
9$^{*}$ & ${\rm H + H^{+}} \rightarrow {\rm H_{2}^{+} + \gamma}$ & L15 \\
10 & ${\rm H_{2}^{+} + H} \rightarrow {\rm H_{2} + H^{+}}$  & A97 \\
11 & ${\rm H_2 + H^+}  \rightarrow {\rm H_2^+ + H} $  & S04 \\
12 & ${\rm H_{2} + e^{-}} \rightarrow {\rm H + H + e^{-}}$ & DS91 \\
13$^{*}$ & ${\rm H_2 + H} \rightarrow {\rm H + H + H}$  & MSM96 \\
14 & ${\rm H^{-} + e^{-}} \rightarrow {\rm H + e^{-} + e^{-}}$  & A97 \\
15 & ${\rm H^{-} + H} \rightarrow {\rm H + H + e^-}$  & A97 \\
16$^{*}$ & ${\rm H^{-} + {\rm H^{+}}} \rightarrow {\rm H + H}$ & CDG99 \\
17 & ${\rm H^{-} + H^{+}} \rightarrow {\rm H_{2}^{+} + e^{-}}$ & A97 \\
18 & ${\rm H_{2}^{+} + e^{-}} \rightarrow {\rm H + H}$ & A97 \\
19 & ${\rm H_{2}^{+} + H^{-}} \rightarrow {\rm H_{2} + H}$ & DL87 \\
20 & ${\rm H + H + H_{2}} \rightarrow {\rm H_{2} + H_{2}}$ & CW83 \\
21 & ${\rm H + H + H} \rightarrow {\rm H_{2} + H}$ & ABN02 \\
22$^{*}$ & ${\rm H^{-} + \gamma} \rightarrow {\rm H + e^{-}}$ & See text \\
23$^{*}$ & ${\rm H_{2}^{+} + \gamma} \rightarrow {\rm H^{+} + H}$ & See text \\
24 & ${\rm H_{2} + \gamma} \rightarrow {\rm H + H}$  & DB96, WG11, See text \\
25$^{+}$ & ${\rm H + H} \rightarrow {\rm H^{+} + e^{-} + H}$  & LCS91 \\
26$^{+}$ & ${\rm H + He}  \rightarrow {\rm H^+ + e^- + He} $ & LCS91  \\
\hline
\end{tabular}
\\ Reactions marked with an asterisk denote cases where we have updated the rate coefficient data compared to the version used in \texttt{Enzo}. The two reactions marked with crosses have been added to \texttt{Grackle} by us and are not included in the original \texttt{Enzo} chemical network described in \citet{2014ApJS..211...19B}. \\
Key: A97 -- \citet{Abel:1997p3456}; ABN02 -- \citet{Abel:2002p131}; BS60 -- \citet{burgess60}; CDG99 -- \citet{croft99}; CW83 -- \citet{cw83}; DB96 -- \citet{Draine:1996p2556}; DL87 -- \citet{dl87}; DS91 -- \citet{ds91}; F92 -- \citet{ferland92}; K10 -- \citet{kreckel10};  L15 -- \citet{Latif15}; LCS91 -- \citet{Lenzuni91}; MSM96 -- \citet{martin96}; S04 -- \citet{savin04a,savin04b}; SLD98 -- \citet{sld98}; WG11 -- \citet{WolcottGreen:2011p121}
\end{table}
\section{Results from IB and CB models}

We plot the results (analogous to Fig.~\ref{fig.conts} and \ref{fig.jhist}) from the IB and CB models in Fig.~\ref{fig.app.conts} and \ref{fig.app.jhist} that qualitatively result in the same conclusions drawn in sec.~\ref{sec.nojcrit}. 

\newpage
\begin{figure*}
\centering
\includegraphics[width=0.75\columnwidth,angle=90,trim={0cm 1cm 1cm 1cm},clip]{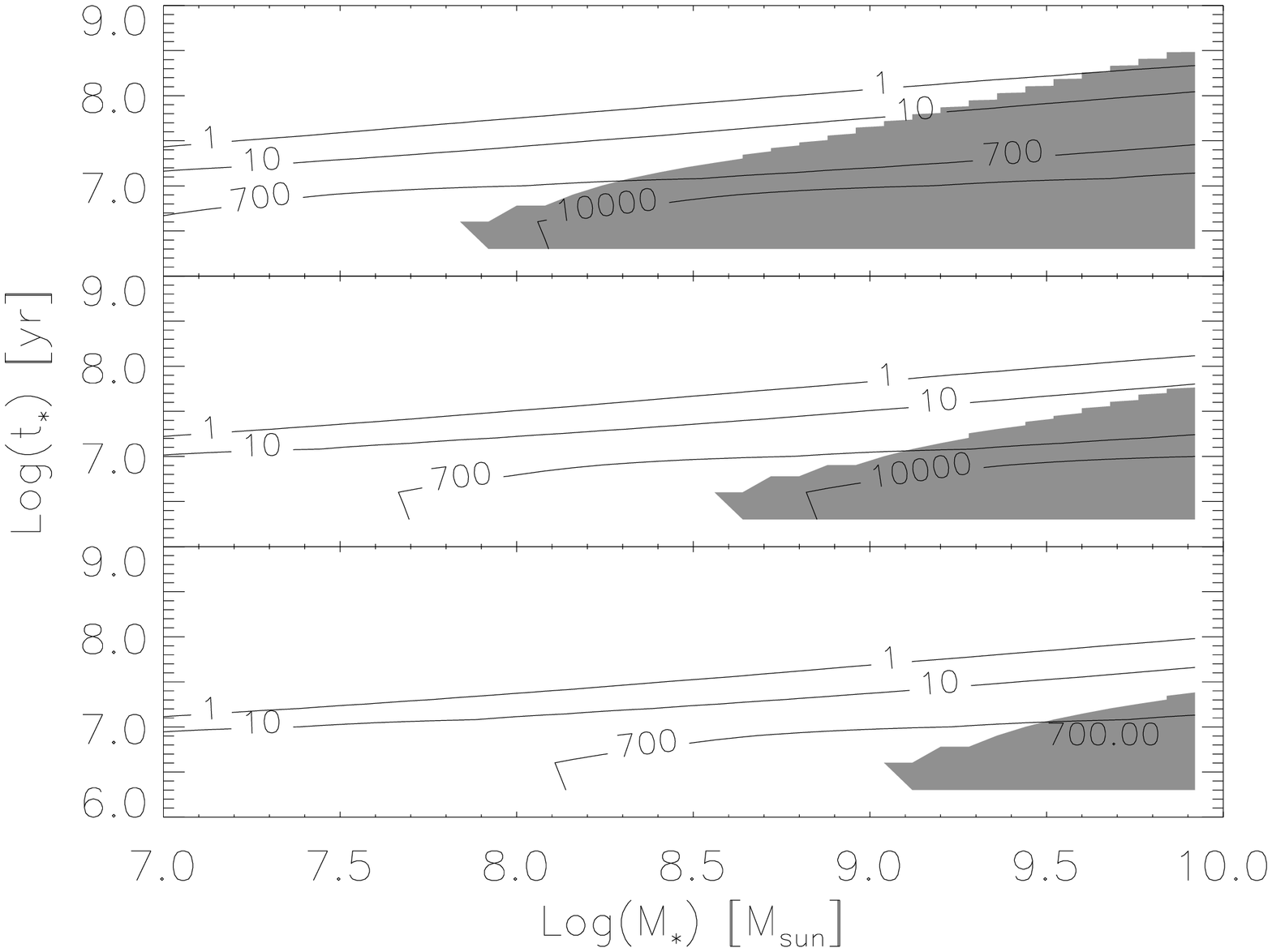}
\includegraphics[width=0.75\columnwidth,angle=90,trim={0cm 1cm 1cm 1cm},clip]{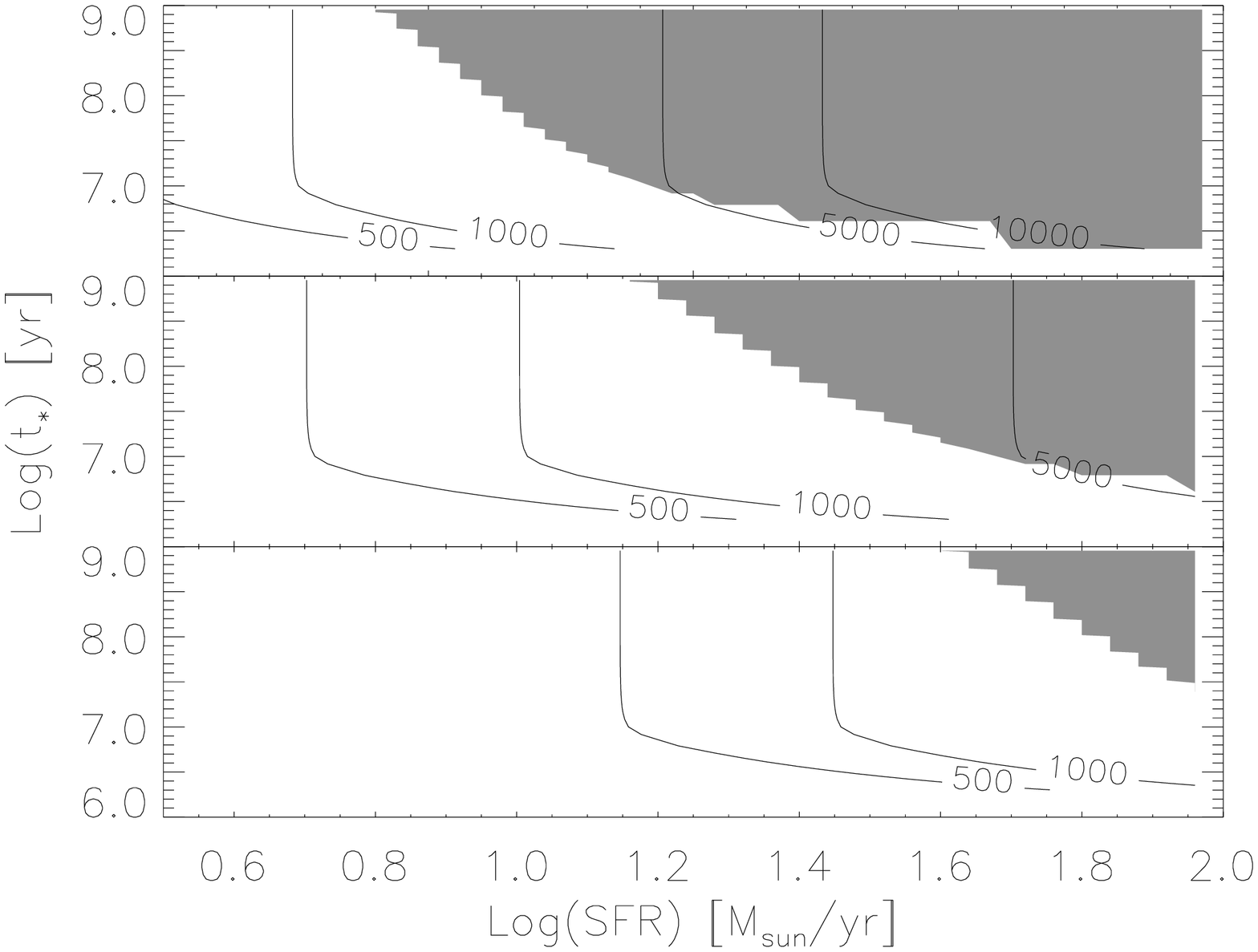}
\caption{Stellar populations that allow for DC. Left: {results for the ISb model, IB}; right: {results for the CSf model, CB}. The shaded regions in grey {indicate the stellar populations} that satisfy Eq. \ref{eq.ratecurve} {for an assumed separation} of 5, 12 and 20 kpc (top, middle and bottom panels, respectively) between the atomic cooling halo and the irradiating source. The contours of $J_{\rm LW}$ at the respective distances are over-plotted in each of the panels.}
\label{fig.app.conts}
\end{figure*}

\begin{figure*}
\centering
\includegraphics[width=0.7\columnwidth,angle=90,trim={-1cm 0cm 0cm -1cm},clip]{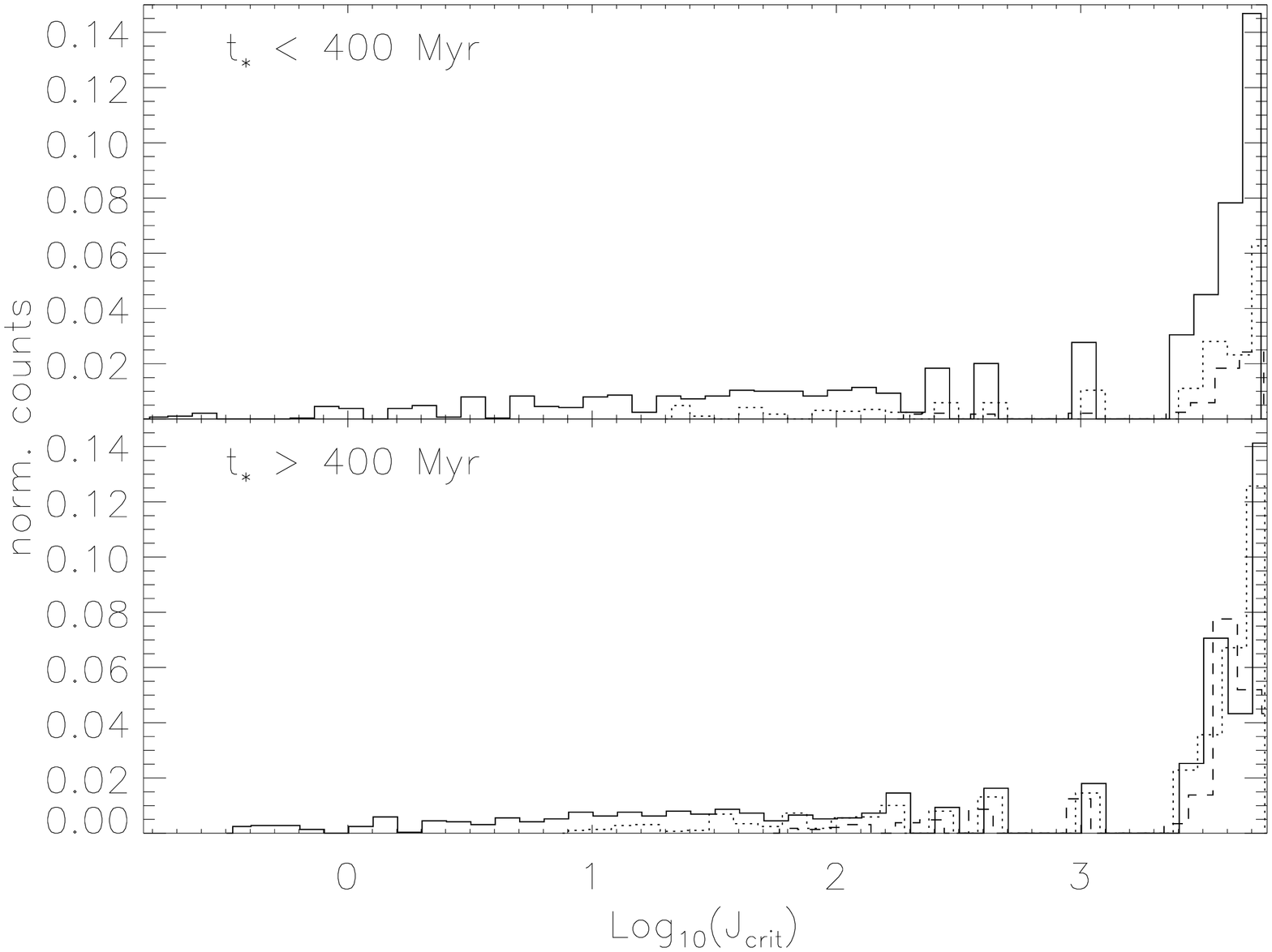}
\includegraphics[width=0.7\columnwidth,angle=90,trim={-1cm 0cm 0cm -0.15cm},clip]{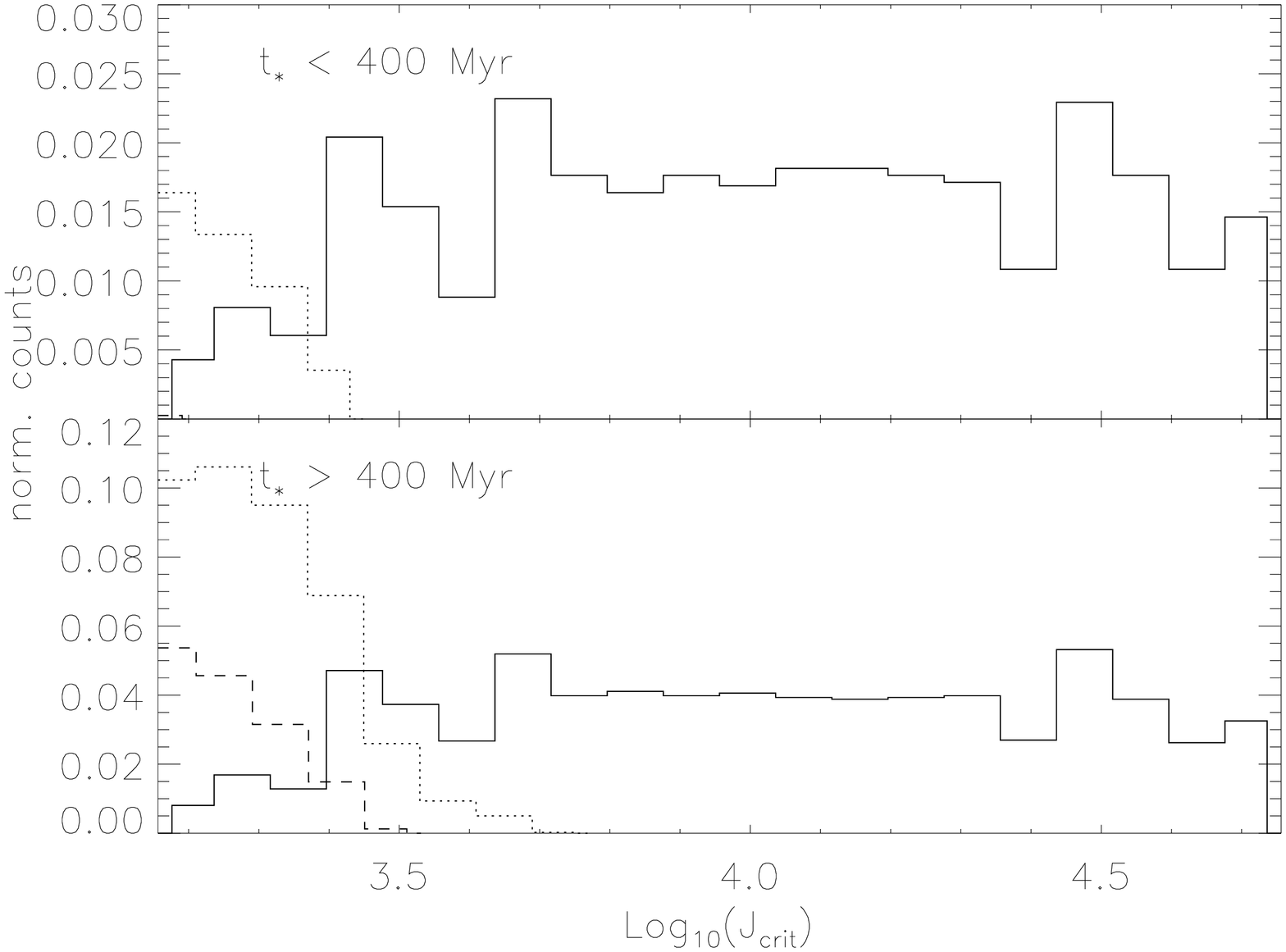}
\caption{Histograms of $J_{\rm crit}$ for the ISb model, IB (left panel) and the CSf model, CB (right panel). The histograms are plotted by splitting the stellar populations into ones with $t_*>400$~Myr and $t_*<400$~Myr. The values of $J_{\rm crit}$ are obtained by requiring that Eq.~\ref{eq.ratecurve} be valid in the grey regions in Fig. \ref{fig.conts}. The solid, dotted and dashed lines correspond to the 5, 12 and 20 kpc separations respectively.}
\label{fig.app.jhist}
\end{figure*}

\label{lastpage}
\end{document}